\documentclass[iop,revtex4,appendixfloats]{emulateapj}

\usepackage{graphicx, rotate}
\usepackage{natbib, lscape}
\usepackage{latexsym, amssymb, verbatim}
\usepackage{eufrak}
\usepackage{amsmath}
\usepackage{epsfig}
\usepackage{times}
\usepackage{rotating}
\usepackage{graphicx}
\usepackage{color}
\usepackage{multirow}

\shorttitle{GMC Collisions as Triggers of Star Formation. V. Observational Signatures}
\shortauthors{Bisbas et al.}

\newcounter{chem}
\newcounter{temp}

\begin{document}

\title{GMC Collisions as Triggers of Star Formation. V. Observational Signatures}

\author{Thomas G. Bisbas\altaffilmark{1,2}, Kei E.~I. Tanaka\altaffilmark{1}, Jonathan C. Tan\altaffilmark{1,3,4}, Benjamin Wu\altaffilmark{4}, and Fumitaka Nakamura\altaffilmark{4}}

\altaffiltext{1}{Department of Astronomy, University of Florida, Gainesville, FL 32611, USA}
\altaffiltext{2}{Max-Planck-Institut f\"ur Extraterrestrische Physik, Giessenbachstrasse 1, D-85748 Garching, Germany}
\altaffiltext{3}{Department of Physics, University of Florida, Gainesville, FL 32611, USA}
\altaffiltext{4}{National Astronomical Observatory of Japan, Mitaka, Tokyo, 181-8588, Japan}
\email{TGB: tbisbas@ufl.edu}

\begin{abstract}

We present calculations of molecular, atomic and ionic line emission
from simulations of giant molecular cloud (GMC) collisions. We
post-process snapshots of the magneto-hydrodynamical simulations
presented in an earlier paper in this series by \citet{Wu17a} of
colliding and non-colliding GMCs. Using photodissociation region (PDR)
chemistry and radiative transfer we calculate the level populations
and emission properties of $^{12}$CO $J=1-0$, [C{\sc i}] $^3{\rm
  P}_1\rightarrow{^3{\rm P}}_0$ at $609\,\mu$m, [C{\sc ii}]
$158\,\mu$m and [O{\sc i}] $^3{\rm P}_1\rightarrow{^3{\rm P}}_0$
transition at $63\,\mu$m. From integrated intensity emission maps and
position-velocity diagrams, we find that fine-structure lines,
particularly the [C{\sc ii}] $158\,\mu$m, can be used as a diagnostic
tracer for cloud-cloud collision activity. These results hold even in
more evolved systems in which the collision signature in molecular
lines has been diminished.
\end{abstract}


\section{Introduction}%

Giant Molecular Cloud (GMC) collisions are a potential mechanism for
triggering star formation activity in such clouds
\citep[e.g.][]{Scov86,Tan00,Higu10,Duar10,Fuku14,Balf15,Tori17}. Models of galactic
shear-driven collisions \citep{Gamm91,Tan00,Task09,Dobb15} may help
explain global galactic star formation relations
\citep{Tan10,Suwa14}. GMC collisions may also be an important
mechanism for driving turbulence within GMCs \citep{Tan13,Jin17,Li17}.
Thus it is important to develop theoretical and numerical models for
such collisions and
then use them to calculate the observational signatures of these
events \citep[e.g.,][]{Duar11,Inou13,Taka14,Hawo15a,Hawo15b}.

With the above goal in mind, this paper continues the works of
\citet{Wu15,Wu17a} (Papers I and II) in studying GMC collisions via
ideal magneto-hydrodynamic (MHD) simulations. In particular, Paper II
studied the kinematics and dynamics of collisions of turbulent,
magnetized, multiphase (via photodissociation region (PDR)-based
heating/cooling functions) GMCs using synthetic observations in the
optically thin limit. Here, we provide a more realistic treatment of
the radiative transfer of line emission from multiple species from
these simulations with a tool we develop that accounts for
velocity-dependent optical depth along the line-of-sight.  Other
papers in this series have added additional physics: \citet{Wu17b}
(Paper III) incorporated star formation via various sub-grid models,
including simulations in which star formation depends on the local
mass to magnetic flux ratio; Christie et al. (2017) (Paper IV) studied
the effects of ambipolar diffusion, especially its effects on the
efficiency of dense core formation. However, here, with our focus on
observational signatures, especially on the larger, global scales of
the GMCs, we return to the simulation outputs of Paper II for our
analysis. We note that these simulations do not include star formation
or any localized feedback, so the results we present will isolate the
``pure'' signature of the GMC-GMC collision process, separate from
these complicating factors.

Our paper is organized as follows. In \S\ref{sec:methods} we give a
brief outline of the snapshots we selected of the MHD simulations
performed in Paper II (\S\ref{ssec:mhd}), of the PDR calculations
(\S\ref{ssec:3dpdr}) and of the radiative transfer tool we develop
(\S\ref{ssec:RT}). In \S\ref{sec:bridge} we discuss the signatures
frequently used by the observational community to identify cloud-cloud
collision activity. In \S\ref{sec:results} we report our results,
which are then further discussed in \S\ref{sec:discussion}. We
conclude in \S\ref{sec:conclusions}.

\section{Numerical methods}
\label{sec:methods}
\subsection{Magneto-hydrodynamical simulations}
\label{ssec:mhd}

For the purposes of this work we use two snapshots from the
three-dimensional MHD simulations performed in Paper II. These
simulations include self-gravity, supersonic turbulence and magnetic
fields (treated in the limit of ideal MHD). They have been performed
with the adaptive-mesh refinement (AMR) code {\sc Enzo}
\citep{Brya14}. 

The two GMCs considered are initially spherical and uniform with total
H-nucleus number density of $n_{\rm H}=100\,{\rm cm}^{-3}$, radius of
$R=20\,{\rm pc}$ and thus with mass of
$M=9.3\times10^4\,M_{\odot}$. As described in Paper II, for
  simplicity the mean particle mass was set to a constant value of
  $\mu=2.33m_{\rm H}$, where $m_{\rm H}$ is the mass of the Hydrogen
  atom. In addition, a constant adiabatic index of $\gamma=5/3$ was
  adopted as the most appropriate single valued choice for a focus on
  the dynamics of the molecular clouds. The ambient medium around the
GMCs consists of gas with $n_{\rm H}=10\,{\rm cm}^{-3}$. The spatial
resolution of the AMR grid has a minimum value of $0.125\,{\rm pc}$. A
magnetic field with strength $B=10\,\mu{\rm G}$ has been included,
directed at an angle of $\theta=60^{\circ}$ with respect to
$x-$direction. The centers of GMCs have an initial separation of $2R$
in the $x-$direction and 0 in the $z-$direction. Along the
$y-$direction they are offset by an impact parameter $b=0.5R$. Both
GMCs have a one-dimensional turbulent velocity dispersion of
$\sigma_v=5.2\,{\rm km}\,{\rm s}^{-1}$.

As described in Paper I, the simulations include PDR-based heating and
cooling processes to determine gas and dust temperature and the
emission properties from various species. We note that these
  heating and cooling functions are applied to both the GMCs and the
  ambient medium, so that the thermal and chemical evolution of the
  gas across the atomic to molecular transition is well-modeled. An
isotropic FUV radiation field with strength $G/G_0=4$ \citep[i.e., 4
  times the][estimate of the local FUV intensity]{Habi68} and a
cosmic-ray ionization rate of $\zeta_{\rm CR}=10^{-16}\,{\rm s}^{-1}$
have been adopted. These values represent conditions observed in
the inner part of the Galaxy at distances of $\sim4\,{\rm kpc}$ from
the Galactic Centre.  Further details are discussed in Papers I and
II.

We select two different cases from the above set of simulations: i)
the case where the clouds collide at a relative speed of $v_{\rm
  rel}=10\,{\rm km}\,{\rm s}^{-1}$ (``colliding'') in which they are
moving at equal but opposite velocities along the axis of collision
(i.e., $+v_{\rm rel}/2=+5\,{\rm km}\,{\rm s}^{-1}$ and $-v_{\rm
  rel}/2=-5\,{\rm km}\,{\rm s}^{-1}$), and ii) the case where the
clouds do not collide (``non-colliding'') but overlap each other along
the line-of-sight, therefore the relative speed is $v_{\rm
  rel}=0\,{\rm km}\,{\rm s}^{-1}$. The selected snapshots for the
colliding case are at times of $t=2$ and $4\,{\rm Myr}$ and are shown
in the top and middle rows of Fig.~\ref{fig:snap}. The collision
occurs along the $x-$direction. For the non-colliding case, we
consider only the snapshot at $t=4\,{\rm Myr}$, which is shown at the
bottom row of Fig.~\ref{fig:snap}.

\subsection{3D-PDR calculations}
\label{ssec:3dpdr}

To perform realistic radiative transfer calculations (see
\S\ref{ssec:RT}) and obtain the emission map of a particular line, we
need knowledge of the corresponding level populations, abundances of
species, as well as gas and dust temperature profiles. To do this, we
use the {\sc 3d-pdr}
code\footnote{https://uclchem.github.io/3dpdr.html} \citep{Bisb12},
which treats the chemistry of PDRs using various cooling and heating
processes. Although the code is able to calculate three-dimensional
PDRs of arbitrary density distributions, the computational cost for
post-processing the above hydrodynamical snapshots at the given
resolution is prohibitively high. Instead, we perform an extended
one-dimensional grid of uniform density slabs irradiated by a
plane-parallel radiation field and we adopt the methodology described
in Paper I, which connects the H-nucleus number density $n_{\rm H}$ of
a cell with a most probable visual extinction value, $A_{V}$.  

The density range spanning the grid of one-dimensional simulations is
$1\,{\rm cm}^{-3}<n_{\rm H}<10^7\,{\rm cm}^{-3}$, with a sampling
every 0.1 dex. This resolution gives a total set of 70 simulations. In
all these calculations we considered a plane-parallel radiation field
with strength $\chi/\chi_0=4$ and a cosmic-ray ionization rate of
$\zeta_{\rm CR}=10^{-16}\,{\rm s}^{-1}$, thus mimicking the conditions
adopted in Papers I and II. We utilized a reduced UMIST 2012 chemical
network \citep{McEl13} of 33 species and 330 reactions and we adopted
the ``standard'' ISM abundances of elements, i.e., [He]/[H]=0.1,
[C]/[H]=$10^{-4}$, and [O]/[H]=$3\times10^{-4}$
\citep{Roel07,Cart04,Card96}.

\begin{figure*}
\center
\includegraphics[width=0.98\textwidth]{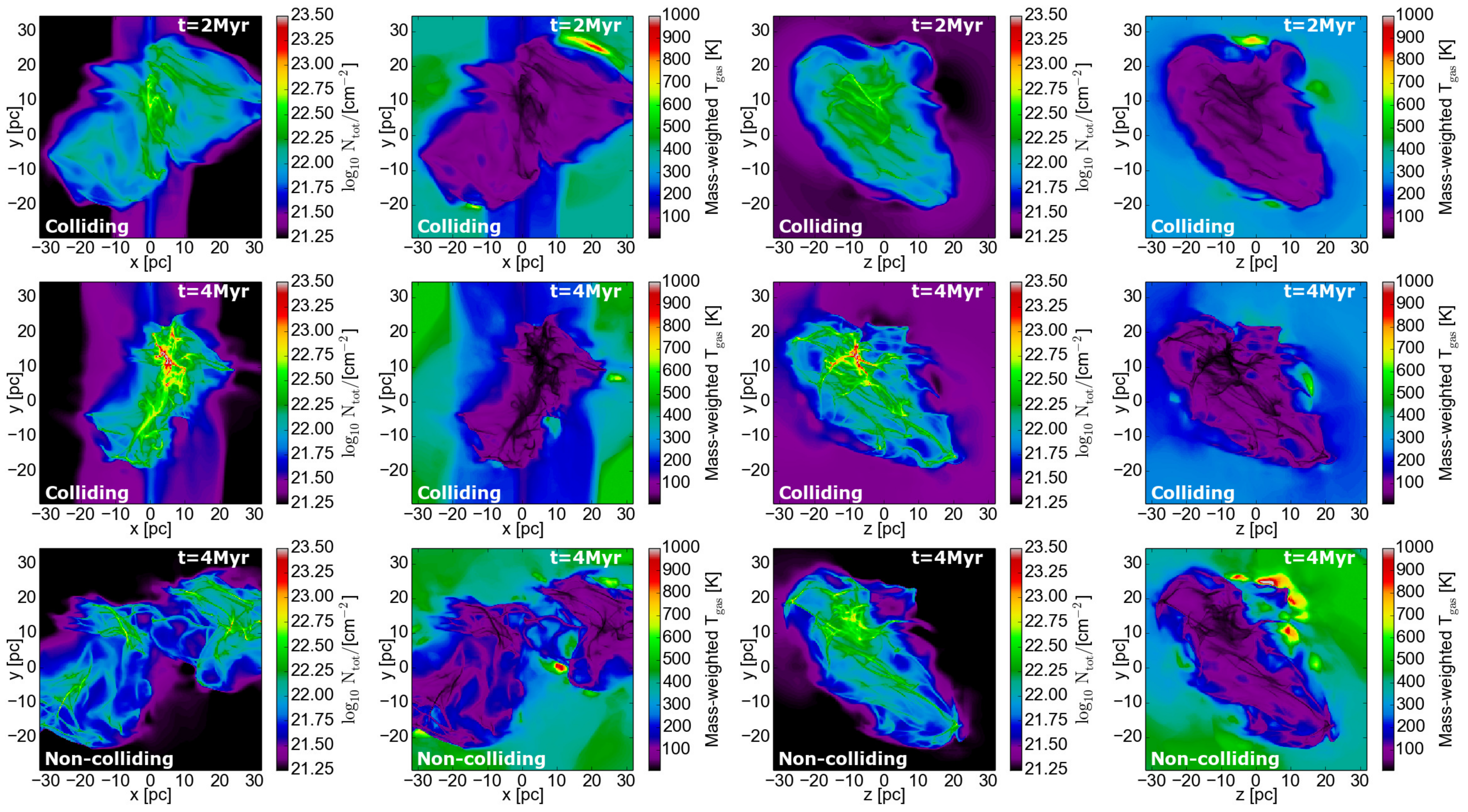}
\caption{
Maps of the total H-nucleus column density (first and third columns)
and mass-weighted gas temperature (second and fourth columns) of the
MHD simulations from Paper II. The integration in the maps of the first
two columns is along the $z-$direction, whereas in the third and
fourth columns it is along the $x-$direction, i.e., the direction
along which the collision occurs. The first and the second rows show
snapshots at $t=2$ and $4\,{\rm Myr}$ of the ``colliding'' case. The
bottom row corresponds to the $t=4\,{\rm Myr}$ snapshot  for the
``non-colliding'' case.}
\label{fig:snap}
\end{figure*}

There is generally good agreement between the grids of 1D simulations
of \citet{Wu15}, which used the simpler Py-PDR code, and this
work. However, in the low density medium that corresponds to low
values of $A_{V}$ we do find some differences in abundances and other
properties, which is not unexpected since PDR codes often deviate in
this regime \citep{Roel07}. This leads to moderate differences in the
calculated mass-weighted gas temperature of Fig.~\ref{fig:snap} in
comparison with \cite{Wu17a}.

\subsection{Radiative Transfer calculations}
\label{ssec:RT}

We construct a radiative transfer tool described in Appendix \ref{app}
which we use for our synthetic observations. As mentioned in
\S\ref{ssec:mhd}, the MHD simulations use an AMR grid, effectively
meaning that the spatial resolution is higher in places of higher
density. For the radiative transfer calculations, however, we convert
the selected snapshots to an uniform grid of 0.125~pc resolution. This
in turn gives a grid consisting of $512^3$ cells which we use to solve
for the radiative transfer along each line-of-sight.

Once the radiative transfer equation is solved (see Appendix
\ref{app}), we calculate the antenna temperature\footnote{Following the notation of \citet{DraiB}}, $T_{A}$, using
the following relation:
\begin{eqnarray}
T_{A}=\frac{c^2I_{\nu}}{2k_{\rm B}\nu^2}.
\end{eqnarray}
Here, $c$ is the speed of light, $k_{\rm B}$ is the Boltzmann
constant, and $I_{\nu}$ is the intensity over frequency
$\nu$, defined as
\begin{eqnarray}
\nu = \nu_0 \left( 1- \frac{v_{\rm los}}{c} \right),
\end{eqnarray}
where $\nu_0$ denotes the frequency in the observer reference frame
and $v_{\rm los}$ is the velocity along the line-of-sight.  The
  integrated antenna temperature over velocity, $W$, is evaluated as
\begin{eqnarray}
W=\int T_A dv_{\rm los},
\end{eqnarray}
where the velocity width adopted to construct the emission maps is
$-15\,{\rm km}\,{\rm s}^{-1}<v_{\rm los}<+15\,{\rm km}\,{\rm
  s}^{-1}$.

Unless otherwise stated, the radiative transfer algorithm has been
applied along the $x-$direction, which is the axis of collision. The
chosen direction of integration is from $+x$ to $-x$, i.e., the
observer is located in the $-x$ direction.
With the collision occuring along the line-of-sight, we expect to see
the clearest signature of its effects in velocity space in this
limiting case.
However, we will also examine how the GMCs appear when viewed from a
direction perpendicular to the collision axis, which is the other
extreme, i.e., for which the signatures of the collision would be
minimized.

\section{The ``bridge-effect'' signature}
\label{sec:bridge}

An observational technique to identify a collision event between two
clouds is to construct their position-velocity (p-v) diagram and look
for two velocity peaks along the line-of-sight.
 Each velocity peak corresponds to one individual cloud.  In the
  p-v diagram, the peaks in $T_{A}$ along the velocity axis at a
  particular position, should be connected via a ``bridge'' of lower
  $T_{A}$. This ``bridge-effect''
indicates the presence of a pair of clouds along the line-of-sight
that are connected by gas at intermediate velocities between the two
main values of each cloud, implying a mutual interaction. 
  Otherwise (i.e., when the velocity peaks are not connected but
  remain isolated), one may simply be seeing two different clouds
  along the line-of-sight that are not interacting with one another.

Bridged velocity peaks have been observed in various cases. For
example, C$^{18}$O $J=1-0$ observations in the Serpens star-forming
region by \citet{Duar10} confirm the existence of a double-peak in the
p-v diagram. The corresponding velocity width has been estimated as
$\sim2\,{\rm km}\,{\rm s}^{-1}$. They explain this feature using
smoothed particle hydrodynamics (SPH) simulations of two colliding
cylinders \citep{Duar11} in which the resulting p-v diagram is
reminiscent to the one observed in the Serpens region. \citet{Tori11}
provide p-v diagrams of the Triffid Nebula (M20) in which a broad
bridge effect connecting multiple velocity peaks along the
line-of-sight is observed. Similarly, \citet{Naka12} presented high
spatial resolution images of the $^{12}$CO $J=1-0$ line for a wide
region including the L1641-N cluster. They find two-velocity component
gas along the line-of-sight suggesting a cloud-cloud collision
event. The velocity width is $\sim3\,{\rm km}\,{\rm s}^{-1}$ which is
about three times higher than the local turbulent velocity ($v_{\rm
  turb}\sim1\,{\rm km}\,{\rm s}^{-1}$).  Recently, \citet{Fuku17a}
identified double velocity peaks in the $^{12}$CO $J=1-0$ line of the
molecular gas toward M42 and M43, which potentially implies that the
northen part of the Orion A cloud may have been formed by two
colliding clouds.

Using hydrodynamical simulations, \citet{Hawo15a,Hawo15b} discuss
extensively how the bridge effect occurs and develops in p-v diagrams
of the $^{12}$CO $J=1-0$ line. They examine various cases and at
different viewing angles of colliding and non-colliding cases and they
additionally examine p-v diagrams of isolated clouds. They also
include modeling of radiative feedback (i.e., H{\sc ii} regions)
generated by stars forming in their simulations. They find that the
bridge feature holds for a small fraction of all possible viewing
angles (e.g., about 20-30\% for a $v_{\rm rel}=10\,{\rm km}\,{\rm
  s}^{-1}$ head-on collision, which is the collision velocity
considered here) and is thus sensitive to the point of view of the
observer.  Furthermore, they find that this signature, when the
geometry is favorable for its detection, is resilient to radiative
feedback from stars formed in the simulations and cannot be easily
reproduced by kinematic behavior other than a cloud-cloud collision. A
bridged double velocity peak can therefore act as a signature of an
undergoing collision event.

Lines other than low-$J$ CO (including isotopologues) have not been
widely considered as diagnostic tracers for GMC-GMC
collisions. However, \citet{Geri15} presented a detection of [C{\sc
    ii}] $158\,\mu{\rm m}$ at position G49.5-0.4 in W51A, in which
strong emission from two velocity components is reported, i.e.,
$T_{A}\sim30\,{\rm K}$ and $\sim25\,{\rm K}$ for velocities of
$\sim55$ and $\sim70\,{\rm km}\,{\rm s}^{-1}$, respectively.  This
implies a relative velocity of $v_{\rm rel}\sim15\,{\rm km}\,{\rm
  s}^{-1}$. W51A is considered to be a strong candidate of GMC-GMC
collision activity \citep{Gins15} and these observations, therefore,
suggest that fine-structure lines may be possible diagnostic
indicators for cloud collision events.

\begin{figure*}
\center
\includegraphics[width=0.97\textwidth]{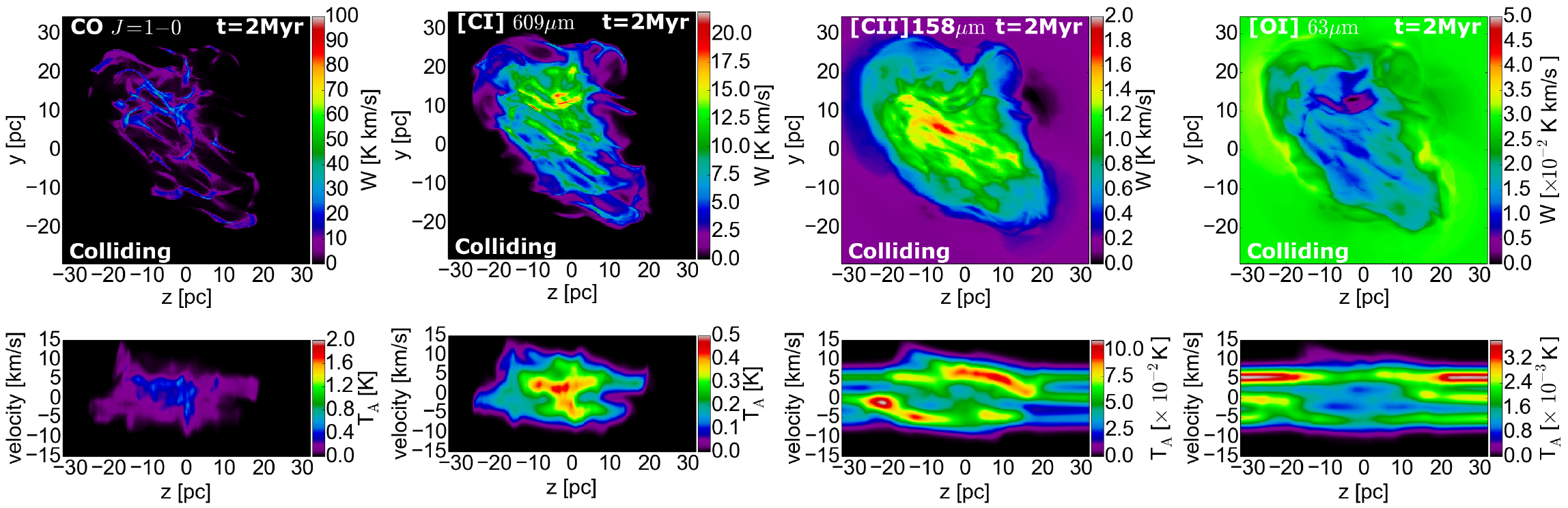}
\includegraphics[width=0.97\textwidth]{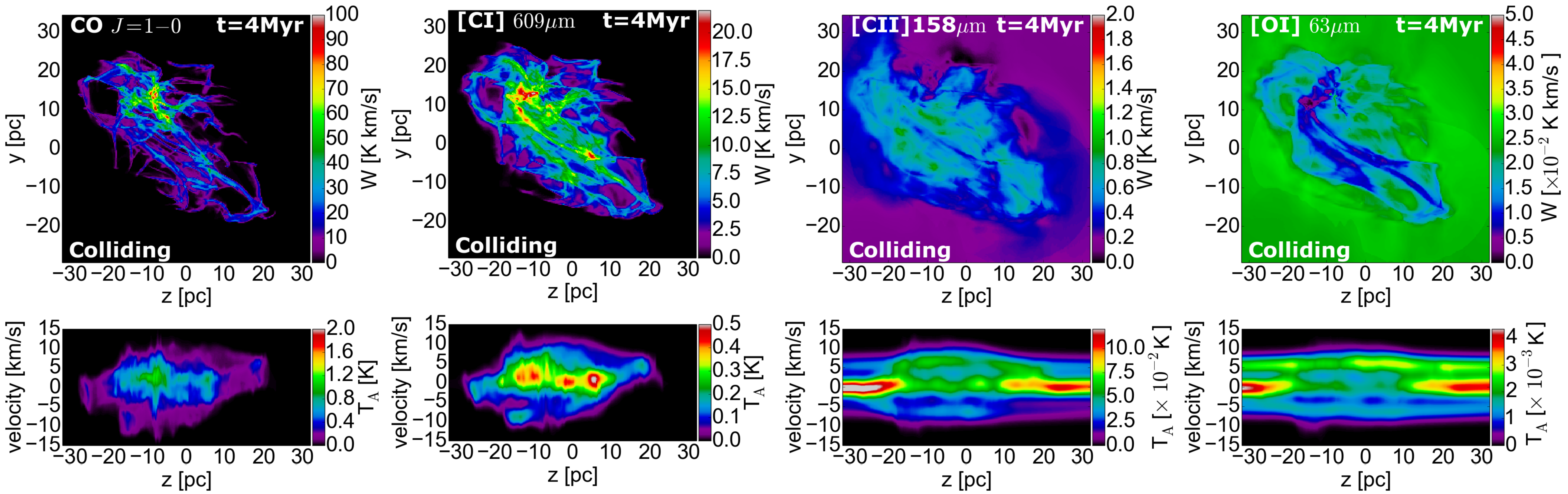}
\includegraphics[width=0.99\textwidth]{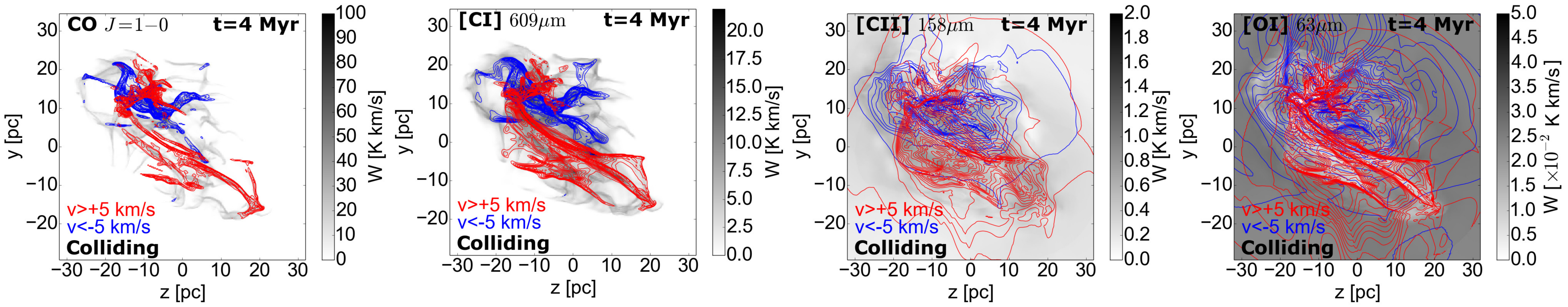}
\caption{
Emission maps (first and third rows) and position-velocity diagrams
(second and fourth rows) for the colliding case at $t=2\,{\rm Myr}$
and $4\,{\rm Myr}$, respectively. The collision occurs along the
line-of-sight. Columns from left to right: CO $J=1-0$; [C{\sc i}]
$609\,\mu{\rm m}$; [C{\sc ii}] $158\,\mu{\rm m}$; O{\sc i}
$63\,\mu{\rm m}$.  The bottom row shows the integrated intensity of
high velocity gas traced by each line at $t=4\,{\rm Myr}$, i.e., maps
the gas that has velocities $>+5\,{\rm km}\,{\rm s}^{-1}$ (red
contours) and $<-5\,{\rm km}\,{\rm s}^{-1}$ (blue contours). The
contours correspond to the intensity of the line. The grayscale shows
the total intensity, $W$, of each line.  In the p-v diagrams and at
$t=2\,{\rm Myr}$, the ``bridge effect'' is better seen in the
fine-structure lines of [C{\sc ii}], [O{\sc i}] and partially in the
[C{\sc i}] line, as all these are primarily emitted from the lower
density gas, including from the ambient atomic medium, that has not
yet undergone collision. Since CO $J=1-0$ is emitted from the
innermost part of both clouds where FUV radiation has been severly
extinguished, identifying the collision signature in this line once
the clouds start to merge is more difficult. Once the collision
further evolves ($t=4\,{\rm Myr}$), the bridge-effect is also
diminished for the [C{\sc i}] $609\,\mu{\rm m}$, although still
holding in [C{\sc ii}] $158\,\mu{\rm m}$ and O{\sc i} $63\,\mu{\rm m}$
lines. The maps of high velocity gas on the other hand show a
collision event at $z\sim-10\,{\rm pc}$ and $y\sim+10\,{\rm pc}$ in CO
$J=1-0$ and [C{\sc i}] $609\,\mu{\rm m}$, with the rest of the
fine-structure lines indicating the bulk movement of GMCs and ambient
medium on larger scales.}
\label{fig:col}
\end{figure*}

\section{Results}
\label{sec:results}

We construct integrated intensity emission maps and p-v diagrams for
all three simulation snapshots considered (two for the colliding case
at $t=2$ and $4\,{\rm Myr}$ and one for the non-colliding case at
$t=4\,{\rm Myr}$). Our standard p-v diagrams are calculated by
averaging over the entire $y-$direction at each different velocity,
$v$, as a function of the $z-$direction. To highlight the bulk motions
that are a signature of a cloud collision, we also show maps of
integrated intensity of only higher velocities, i.e., $>\!+5\,{\rm
  km}\,{\rm s}^{-1}$ (redshifted) and $<\!-5\,{\rm km}\,{\rm s}^{-1}$
(blueshifted). This emission will mostly correspond to gas that moves
at velocities greater than those reached due to internal turbulence.
We focus our consideration on four different lines: $^{12}$CO $J=1-0$
(hereafter CO $J=1-0$); [C{\sc i}] ($609\,\mu$m); [C{\sc ii}]
$158\,\mu$m; and [O{\sc i}] $63\,\mu$m.

\subsection{Colliding case}
\label{ssec:colliding}

\subsubsection{Fiducial results: comparison of CO and atomic fine structure lines}

Figure \ref{fig:col} shows the emission maps (integrated intensity,
$W$) and the corresponding p-v diagrams for the colliding case at
$t=2$ and $4\,{\rm Myr}$ for all different lines considered. The
bottom row shows the high velocity gas emission for each case, which
we discuss further below. These figures illustrate how each different
species reveals different parts of the clouds and their surroundings.

The CO $J=1-0$ transition, which has the strongest emission,
originates from the densest parts of the GMCs. Since the collision
occurs along the line-of-sight and the density increases as time
progresses, at $t=4\,{\rm Myr}$ the intensity of this CO transition is
enhanced compared to that at 2~Myr. The enhancement is more prominent
in the particular region located at $z\sim-10\,{\rm pc}$ and
$y\sim+10\,{\rm pc}$, where the collision mainly occurs. Here, the
total H-nucleus column density reaches a value of
$\gtrsim2\times10^{23}\,{\rm cm}^{-2}$.  The enhancement in this line
is partly because the higher density of the colliding region rapidly
extinguishes the ambient FUV radiation field.

In the p-v diagrams, while the two velocity components of the original
GMCs are just about discernible at 2~Myr, overall, and especially by
4~Myr, the emission becomes dominated by gas that has been compressed
in the collision, i.e., near zero velocity in the simulation
frame. Thus a bridge-effect signature linking two distinct peaks is
not a characteristic feature of the CO $J=1-0$ emission at these
times.

We performed two additional investigations related to the molecular
gas emission. First, we examined the equivalent p-v diagrams of CO
$J=3-2$, which originates from relatively dense molecular gas. We
found these p-v diagrams exhibit very similar morpholigies to those of
CO $J=1-0$. Second, we constructed the p-v diagram of the CO $J=1-0$
at $t=1\,{\rm Myr}$, i.e., at an earlier stage of the evolution. We
found that at this stage, the bridge-effect signature linking two
distinct peaks was clearly visible, although its brightness
temperature was much weaker than that shown in Fig.~\ref{fig:col},
since relatively fewer higher-density clumps were present at these
earlier times.

\begin{figure*}
\center 
\includegraphics[width=0.95\textwidth]{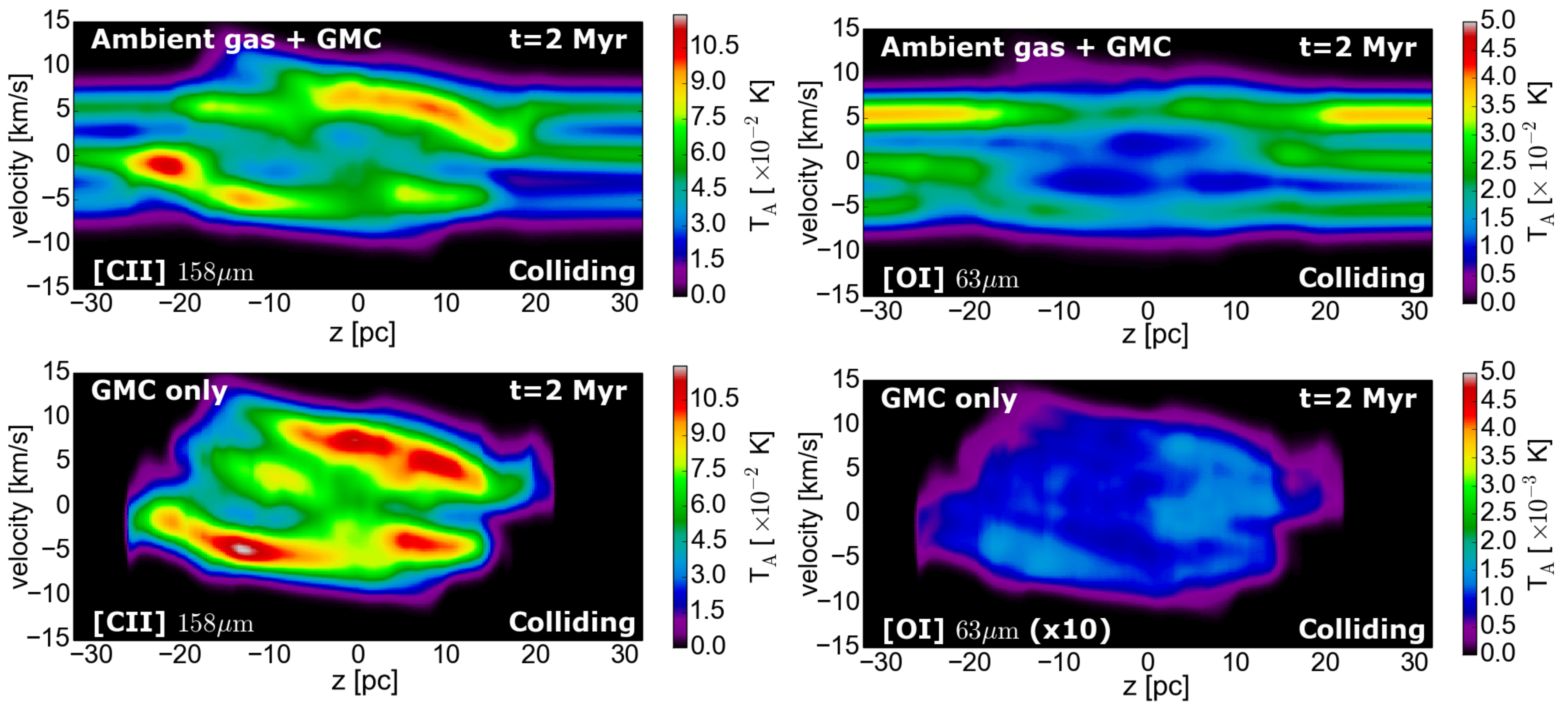}
\caption{
Position-velocity diagrams of [C{\sc ii}] $158\,\mu$m (left column)
and [O{\sc i}] $63\,\mu$m (right column) for the colliding case at
$t=2\,{\rm Myr}$.  In the top row, we include the contributions from
both the ambient ISM and the GMC (i.e., same as in
Fig.~\ref{fig:col}). In the bottom panel, we show only the
contribution of the GMCs, i.e., selecting emission only from regions
with $n_{\rm H}>20\:{\rm cm}^{-3}$ (see \S\ref{ssec:ambient}). We find that [C{\sc
    ii}] $158\,\mu$m emission is dominated by that from the GMCs,
rather than the ambient medium. Indeed, the brightness temperature of
the [C{\sc ii}] $158\,\mu$m line emitted from the GMC-only case is
stronger than the corresponding one from the total case, due to
optical depth effects. However, the [O{\sc i}] $63\,\mu$m
line emitted from the GMCs is significantly weaker than that from the
ambient medium (note that we enhance the brightness temperature by a
factor of 10 in the lower panel, to match the color-bar range of the
upper panel). As such, the bridge-effect of GMC material is seen only
in the [C{\sc ii}] line and not in [O{\sc i}].  }
\label{fig:pvoi}
\end{figure*}

\begin{figure}
\center 
\includegraphics[width=0.45\textwidth]{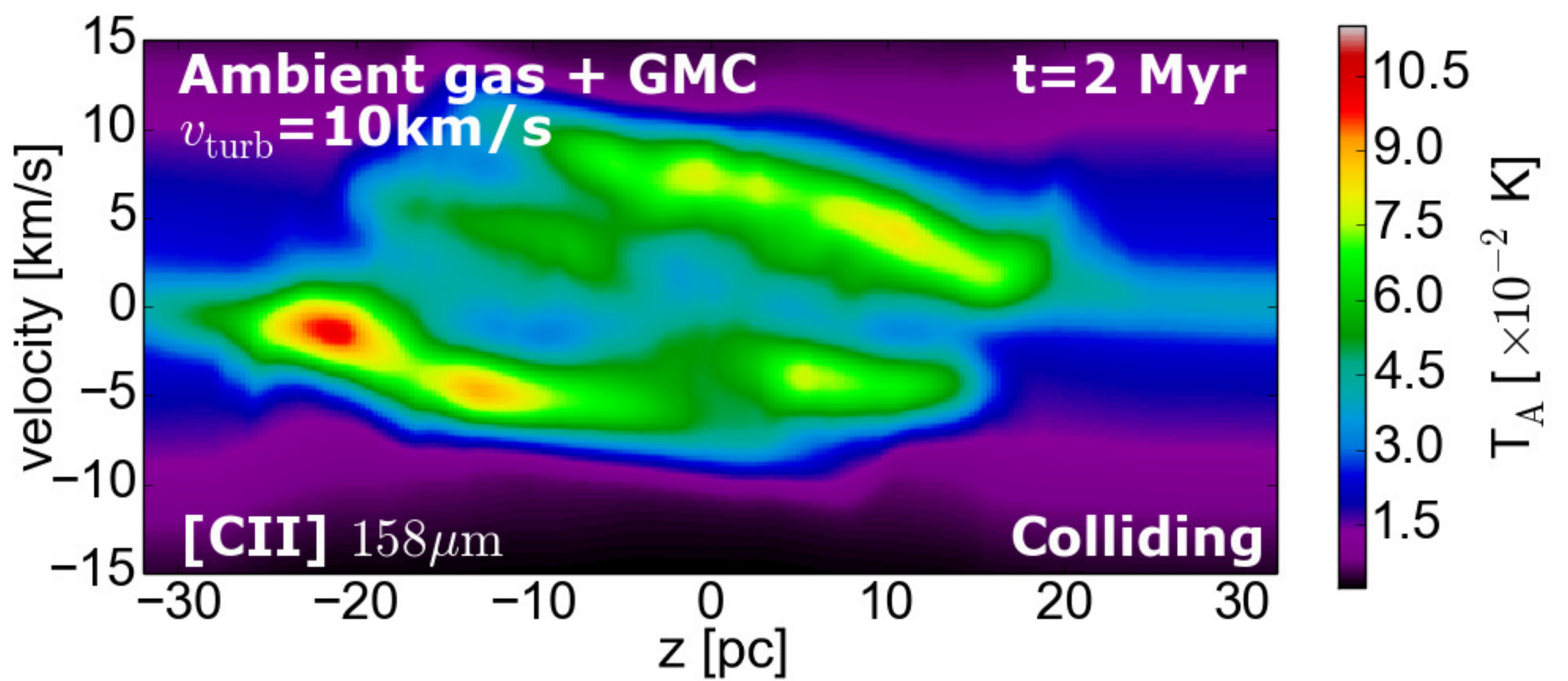}
\caption{
Position-velocity diagram of [C{\sc ii}] $158\mu$m for the
colliding case at $t=2$. Here, we have performed radiative transfer
calculations considering a turbulent velocity of $v_{\rm
  turb}=10\,{\rm km}\,{\rm s}^{-1}$ in the ambient medium (see
\S\ref{ssec:ambient}). Even in the case of an ambient medium with
enhanced turbulent velocity dispersion, the bridge-effect in [C{\sc
    ii}] $158\mu$m emission can still be clearly observed.  }
\label{fig:geo}
\end{figure}

\begin{figure}
\center 
\includegraphics[width=0.47\textwidth]{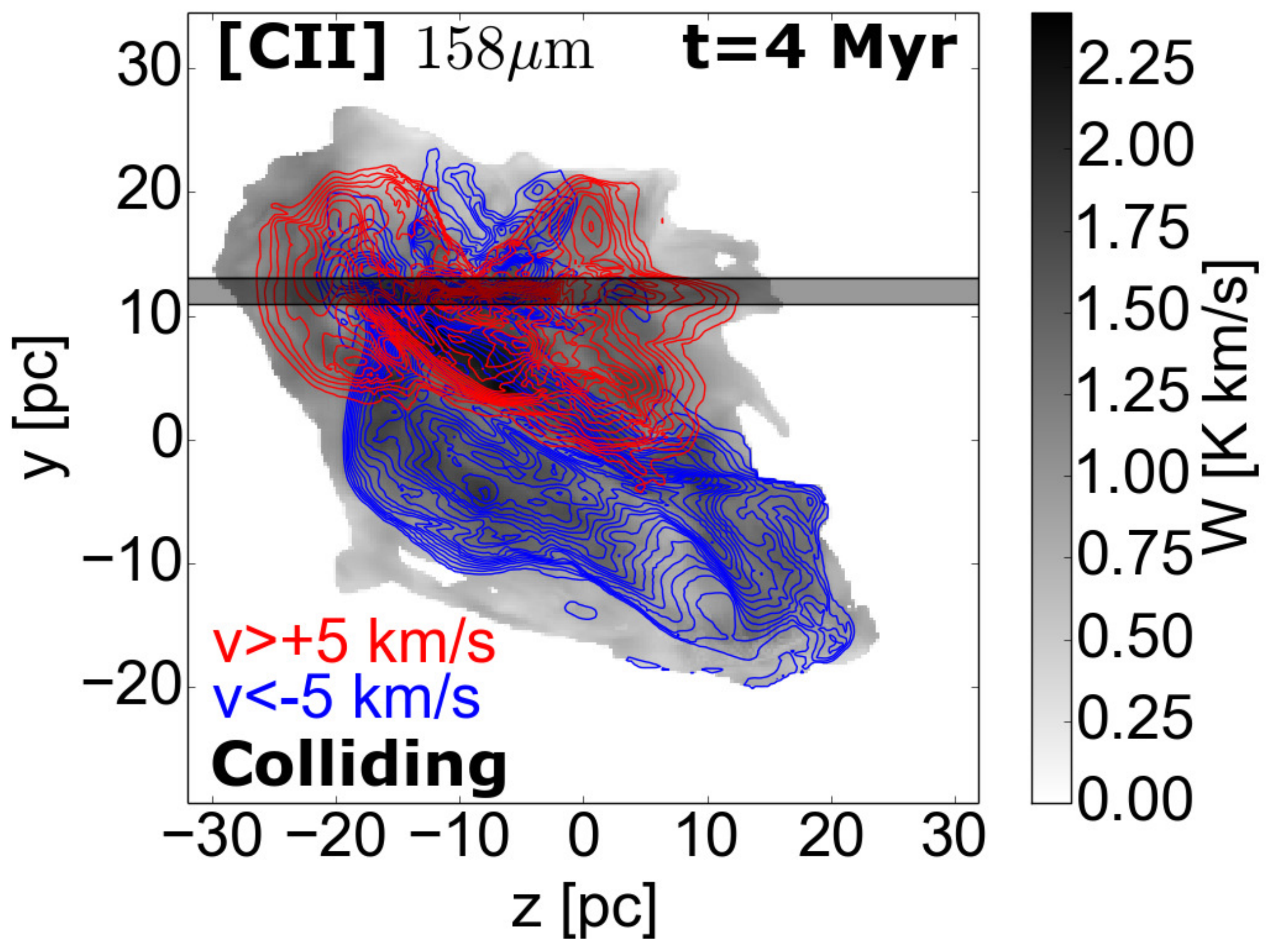}
\includegraphics[width=0.47\textwidth]{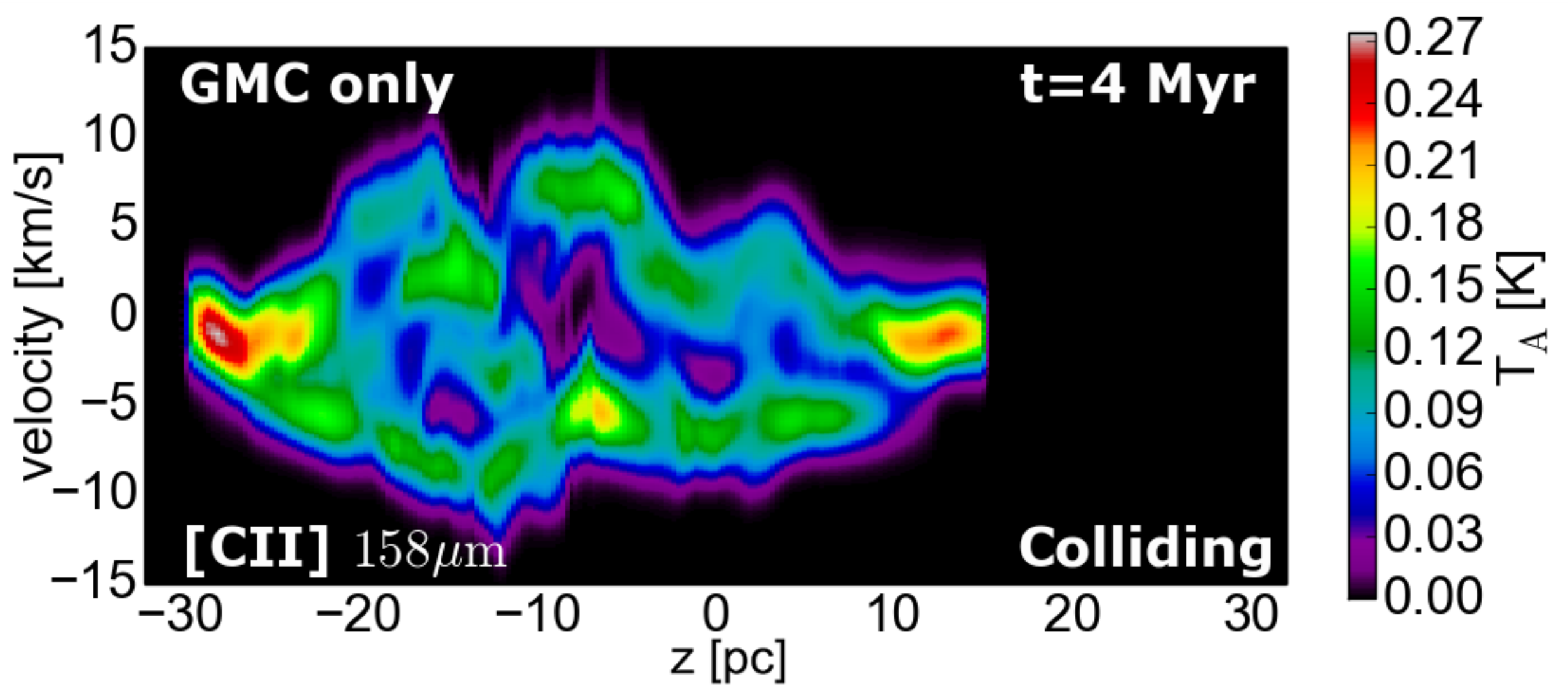}
\caption{ 
{\it Top panel:} High velocity gas in the [C{\sc ii}]
$158\,\mu$m line at $t=4\,{\rm Myr}$ when the contribution of the ISM
has not been considered (GMC only case). It can be seen that the two
red and blue shifted components overlap, indicating occurence of a
cloud collision. The shadowed strip of 2 pc width, shows the part of
the GMCs whose p-v diagram is shown in the bottom panel.
{\it Bottom panel:} p-v diagram extracted from the 2-pc-wide strip
through the dense overlap region. The high-velocity peaks in the upper
panel are bridged indicating the collision is traced by the [C{\sc
    ii}] line. 
}
\label{fig:cuts}
\end{figure}

The [C{\sc i}] $609\,\mu{\rm m}$ line is primarily emitted from the
more diffuse gas of both GMCs and hence it reveals the majority of the
molecular gas of the clouds
\citep[see also][]{Papa04,Offn14,Bisb15,Bisb17,Glov16}. Note that its intensity
remains approximately constant as time increases from $t=2$ to
$t=4\,{\rm Myr}$.

The fine-structure line of [C{\sc ii}] $158\,\mu$m is emitted from the
outer parts of both GMCs, since it is produced by the interaction with
the isotropic FUV radiation field. This line is also emitted from the
surrounding ambient medium, i.e., the ISM gas 
that is here set up as an idealized representation of atomic cold
neutral medium with initial $n_{\rm H}=10\:{\rm cm}^{-3}$, but also
moving alongside each GMC and so forming a shock-compressed layer near
zero velocity (see Paper II). The relative importance of the emission
from the ambient medium compared to that from the GMCs is discussed
below.
In Fig.~\ref{fig:col} we notice that
the [C{\sc ii}] emission decreases in intensity from 2 to $4\,{\rm
  Myr}$, which reflects the conversion of more material from the
clouds in CO-emitting structures.

Similarily to [C{\sc ii}] $158\,\mu$m, the [O{\sc i}] $63\,\mu$m line,
although much weaker than all the above lines, originates from the
outermost parts of both clouds and with significant contributions from
the ambient medium.
The $63\,\mu$m line is optically thick and due to this, its intensity
is noticeably reduced at the parts of the map that correspond to high
column densities (see Fig.~\ref{fig:snap}). The blue-shifted component
in the p-v diagrams (i.e., for $v_{\rm los}\sim-5\,{\rm km}\,{\rm
  s}^{-1}$) has a reduced antenna temperature compared to the
red-shifted component.  The three well-defined stripes of antenna
temperature correspond to the contribution of the ambient medium. The
upper and lower stripes are due to the red-shifted and blue-shifted
component respectively (moving away and towards the observer at a
constant speed of $v_{\rm rel}\pm5\,{\rm km}\,{\rm s}^{-1}$). 
The central strip is mostly due to the shock compressed ambient
medium (see \S\ref{ssec:ambient}).
We further note that the [O{\sc i}] line and in particular its ratio
with [C{\sc i}] $609\,\mu{\rm m}$ can be used as diagnostic of the
intensity of the FUV radiation in PDRs \citep[see][]{Bisb14}.

Summarizing the above results, the p-v diagrams in all cases span a
width in velocity space $\sim\pm10\,{\rm km}\,{\rm s}^{-1}$, i.e.,
resulting from the relative motion of both clouds, that is larger than
that due to internal turbulence within the initial GMCs, i.e., $\sim
\pm5\,{\rm km}\,{\rm s}^{-1}$. Once the collision process begins,
$W_{\rm CO(1-0)}$ becomes enhanced. The motion of the gas then becomes
more turbulent due to the collision as the two GMCs merge. This is
reflected in the p-v diagram at $t=4\,{\rm Myr}$ in which the overall
width in velocity space has increased. As a result, the double
velocity peak feature that could indicate collision is diminished
\citep[see also][]{Hawo15b}, although at the position $\sim-10\,{\rm
  pc}$ (where the collision occurs) the intensity is quite strong over
most of the velocity width of $\sim20\,{\rm km}\,{\rm s}^{-1}$.  The
p-v diagram of the [C{\sc i}] $609\,\mu{\rm m}$ line at $t=2\,{\rm
  Myr}$ indicates, however, that such a feature is reminiscent of the
bridge-effect discussed in \S\ref{sec:bridge} and in particular at
position $\sim0\,{\rm pc}$.  As time progresses and the clouds merge
($t=4\,{\rm Myr}$), this signature disappears.

For systems whose collision is evolved to the point of erasing or
diminishing any potential bridge-effect feature in the low-$J$ CO p-v
diagram, fine-structure lines may be promising alternatives to be
diagnostics for such activity. This can be seen in the p-v diagrams of
[C{\sc ii}] $158\,\mu$m and [O{\sc i}] $63\,\mu$m; these two lines
give the most prominent signature of cloud-cloud collision, as the
bridge effect linking two distinct peaks can now be clearly
seen. However, as discussed above, the emission of both [C{\sc ii}]
$158\,\mu$m and [O{\sc i}] $63\,\mu$m lines originate mostly from the
outer envelope of both GMCs and the surrounding ambient ISM gas, some
of which corresponds to gas that has not yet suffered any collision
and still carries the information of the relative velocity of GMCs
prior to collision. They can thus be used in order to identify and/or
clarify the collision process in systems whose low-$J$ CO lines may
provide limited information.

The bottom row of Fig.~\ref{fig:col} shows the high velocity gas
traced by each line. These maps correspond to the $t=4\,{\rm Myr}$
snapshots. The grayscale shows the total intensity, $W$, of each
line. Red contours correspond to redshifted gas with velocities
$v>+5\,{\rm km}\,{\rm s}^{-1}$ and blue contours to blueshifted gas
with velocities $v<-5\,{\rm km}\,{\rm s}^{-1}$. In both cases, the
contours correspond to the intensity of the line. We do not include
velocities in the range $-5<v<+5\,{\rm km}\,{\rm s}^{-1}$ as we want
to isolate the bulk motion of the gas due to collision and the mutual
interaction of the GMCs rather than additionally mapping the width
reached due to turbulence. The Doppler shifts of both the CO $J=1-0$
transition and the [C{\sc i}] $609\,\mu{\rm m}$ line reveal the
collision at $z\sim-10\,{\rm pc}$ and $y\sim+10\,{\rm pc}$
corresponding to the region with the highest density. However,
  the fine-structure line of [C{\sc ii}] $158\,\mu$m reveals the bulk
  movement of the GMCs and is thus a good indicator of the cloud-cloud
  collision.

\subsubsection{Contribution of the ambient medium and the utility of the $158\,\mu$m line
as a cloud-cloud collision diagnostic}
\label{ssec:ambient}

Emission from the ambient, lower density medium can become important
when studying the above fine-structure lines in these simulations. To
clarify the contribution of this gas in our results, we calculate the
p-v diagrams of [C{\sc ii}] $158\,\mu$m and [O{\sc i}] $63\,\mu$m
lines emitted by densities $n_{\rm H}>20\,{\rm cm}^{-3}$ for the case
at $t=2\,{\rm Myr}$.  For densities below $20\,{\rm cm}^{-3}$, we
consider negligible contribution in optical depth. This ensures that
the p-v diagram will contain information that is essentially only from
the GMCs. These results are shown in Fig.~\ref{fig:pvoi}: the upper
row shows the total emission and the bottom row shows the ``GMC-only''
emission. The bridge-effect is seen in the p-v diagram of [C{\sc ii}]
$158\,\mu$m, as it is primarily emitted from the outer envelopes of
the GMCs. Indeed, for our particular set-up the GMC-only emission
appears stronger since in the total case absorption from the ambient
medium acts to diminish the signature of the collision. This may
indicate an important role for [$^{13}$C{\sc ii}] observations to
complement [C{\sc ii}] studies \citep[e.g.][]{Goic15}.

On the other hand, [O{\sc i}] $63\,\mu$m is mainly emitted from the
ambient medium and removing the contribution of the latter from our
calculations results in a very weak brightness temperature (note that
$T_{A}$ in the bottom right panel of Fig.~\ref{fig:pvoi} is
enhanced by a factor of 10).  We thus find that the fine-structure
line of [O{\sc i}] $63\,\mu$m, being ambient medium dominated, is not,
on its own, a good tracer of GMC-GMC collisions, but rather can give
information on colliding atomic flows. However, since molecular clouds
are known to have significant atomic halos \citep[e.g.][]{Ande92,Wann99}
[O{\sc i}] observations may still provide important complementary
information for understanding particular collision candidates.

By design, the treatment of the ambient atomic medium in the Wu et
al.~(2017a) simulations is highly idealized. In particular, no
internal turbulence is included in this component. Thus to explore
potential effects of such turbulence, for the case of the [C{\sc ii}]
$158\,\mu$m line emission, we perform an additional radiative transfer
calculation in which we assume an enhanced turbulent velocity (in
post-processing) of $v_{\rm turb}=10\,{\rm km}\,{\rm s}^{-1}$ (see
Eqn.~\ref{eqn:sigma}) for densities below $20\,{\rm cm}^{-3}$.
Figure~\ref{fig:geo} shows the p-v diagram of this test. Comparing
this with the corresponding figure in Fig.~\ref{fig:col} (top left
panel), we see that the velocity width is now increased and that the
brightness temperature is moderately decreased. This is due to
increased optical depth from the ambient medium over the range of
velocities of [C{\sc ii}] emission from the GMCs.
Still, we find that the bridge-effect is resilient to the
uncertainties introduced from the emission of the ambient medium and
the [C{\sc ii}] $158\,\mu$m line is thus an important and useful
diagnostic for GMC-GMC collisions.

The utility of [C{\sc ii}] $158\,\mu$m as a diagnostic of cloud-cloud
collisions is further demonstrated in Fig.~\ref{fig:cuts}, in which
the upper panel shows a map of the high-velocity gas at $t=4\,{\rm
  Myr}$ and where we exclude the contribution of the ambient
medium. Extended parts of the high-velocity gas overlap, particularly
in the region corresponding to the densest part of the colliding GMCs
and where star formation is likely to be triggered (Paper III). We
then isolate this part of emission map to construct a p-v diagram
corresponding to a cut of only 2 pc width.
This p-v diagram, also shown in Fig.~\ref{fig:cuts}, reveals that the
emission peaks at distinct velocities are bridged in p-v space.

\subsubsection{Effect of viewing angle - limiting case of a side-on view}

\begin{figure}
\center
\includegraphics[width=0.37\textwidth]{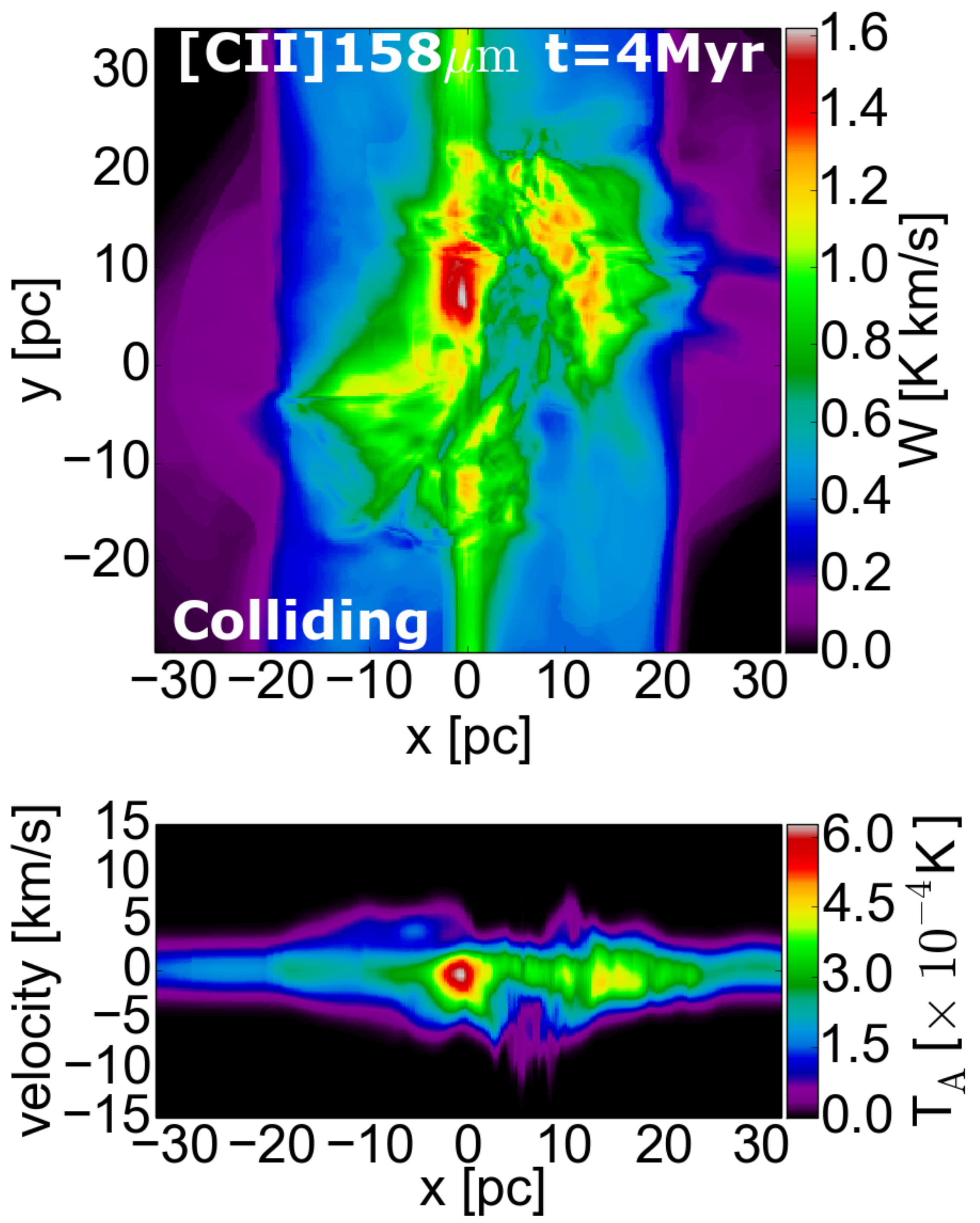}
\caption{
Emission map and p-v diagram of the colliding case in the [C{\sc ii}]
$158\mu$m line at $t=4\,{\rm Myr}$. The radiative transfer
calculations have been performed along the $z-$direction which is
perpendicular to the direction of the collision. The stripe at
$x\sim0\,{\rm pc}$ and along the $y-$direction in the emission map is
due to the collision of the ambient medium. As expected, in the p-v
diagram we do not observe any feature pointing to a collision
event. Instead, the width in velocity space is due to the internal gas
motions (i.e., internal turbulence) rather than the bulk motion of
each GMC.}
\label{fig:xy}
\end{figure}

In Fig.~\ref{fig:xy} we show the emission map and p-v diagram of the
[C{\sc ii}] line at $t=4\,{\rm Myr}$ when the observation is made
perpendicular to the direction of the collision.  In this case, any
signature of the cloud-cloud collision is minimized, and the velocity
width now is in principle connected with the internal turbulence of
the gas. However, the emission map shows a characteristic feature of a
rectangular shape spanning $\sim40\,{\rm pc}$ (from $-20$ to
$+20\,{\rm pc}$) with a bright stripe at $x\sim0\,{\rm pc}$ across the
$y-$axis. This originates from the ambient ISM gas in which both GMCs
are embedded. The p-v diagram shows a single, main component in
velocity space. Still, in the typical case one expects some part of
the collision axis to be along the line of sight and the purely
perpendicular case will be a relatively unlikely occurence.

\subsection{Non-colliding case}
\label{ssec:noncolliding}

\begin{figure*}
\center
\includegraphics[width=0.97\textwidth]{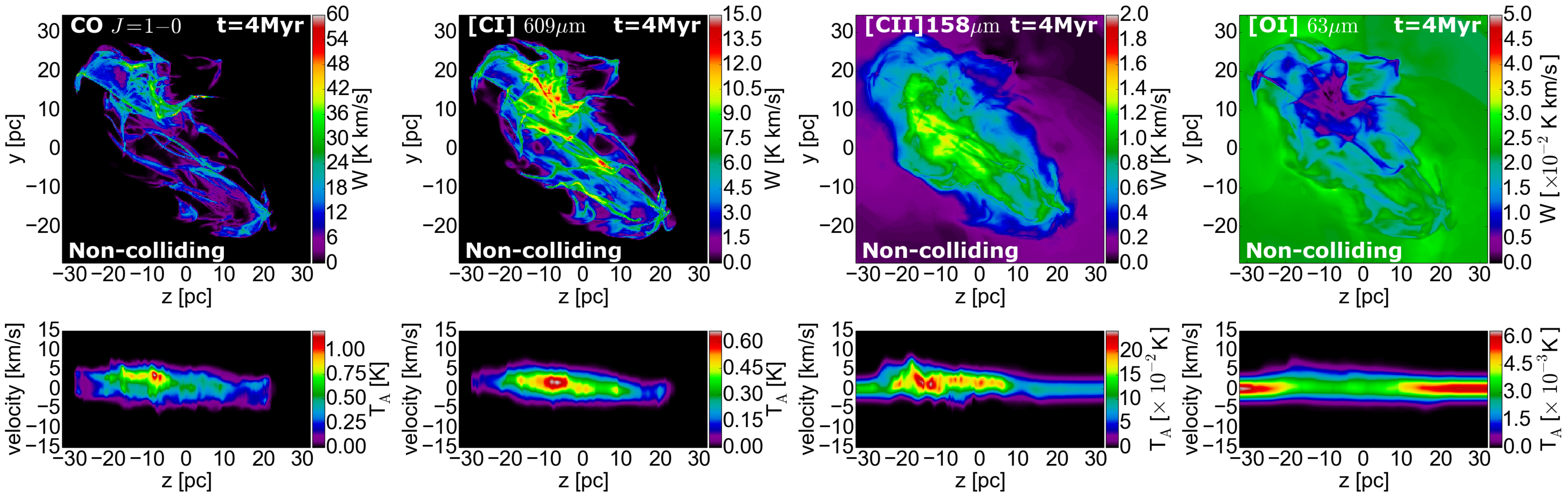}
\includegraphics[width=0.99\textwidth]{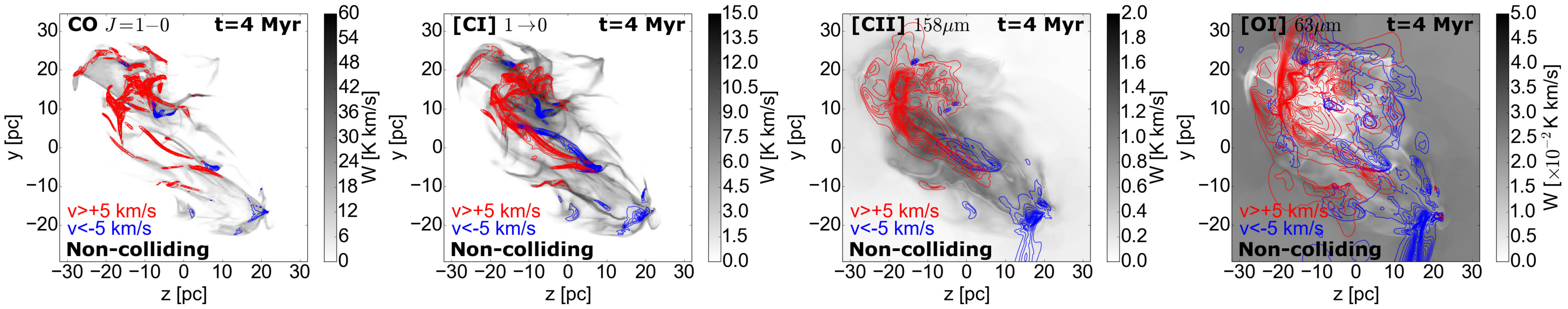}
\caption{
As in Fig.~\ref{fig:col}, but now for the non-colliding case. Here we
plot the emission maps, p-v diagrams and the high velocity gas for
$t=4\,{\rm Myr}$ only. From the p-v diagrams it can be seen that in
all cases we find a width of $<8\,{\rm km}\,{\rm s}^{-1}$
corresponding to the turbulent velocity dispersion due to the internal
gas motions in each GMC. However, the gravitational forces acting on
both clouds result in their mutual attraction. This can be observed in
the high velocity gas maps in which the upper-left part is redshifted
and the bottom-right is blueshifted. Still, in none of the cases do we
observe an interaction pointing to a collision, such as seen in
Fig.~\ref{fig:col}. 
}
\label{fig:nocol}
\end{figure*}

Figure~\ref{fig:nocol} shows the emission maps, p-v diagrams and high
velocity gas maps for the non-colliding case at $t=4\,{\rm Myr}$. The
emission maps in all four different lines are qualitatively similar to
the colliding case at this dynamical time (see Fig.~\ref{fig:col} for
comparison). However, the intensities of CO $J=1-0$ and [C{\sc i}]
$609\,\mu{\rm m}$ are now lower since the system does not reach the
necessary high densities and therefore the high column densities as in
the colliding case. On the other hand, the emission maps of [C{\sc
    ii}] and [O{\sc i}] lines do not significantly differ from the
colliding case. The lower column densities reached in this simulation
do not shield the $158\,\mu$m line and it thus remains brighter than
the corresponding one of Fig.~\ref{fig:col}.

The velocity widths of the p-v diagrams in Fig.~\ref{fig:nocol} are
all within the $-5\lesssim v_{\rm los}\lesssim+5\,{\rm km}\,{\rm
  s}^{-1}$ range which results from the turbulent internal motion of
the gas in both GMCs. An interesting feature, however, is that the
intensity-weighted velocity gradients for the CO $J=1-0$ and [C{\sc
    i}] $609\,\mu{\rm m}$ lines are decreasing along the position axis
in the p-v diagrams. This is due to the mutual gravitational forces
attracting the GMCs towards each other in the simulation domain. These
gradients are most clearly seen in the high velocity gas maps in the
bottom row of the figure. Here it can be seen that the red- ($v_{\rm
  los}>+5\,{\rm km}\,{\rm s}^{-1}$) and blue- ($v_{\rm los}<-5\,{\rm
  km}\,{\rm s}^{-1}$) shifted components do not overlap as shown in
the bottom row of Fig.~\ref{fig:col}, different from the cloud-cloud
collision case. For example, for the particular geometries
  considered we find a $\lesssim17\%$ overlap in the [C{\sc ii}]
  $158\,\mu$m line as opposing to a $\gtrsim42\%$ in the colliding
  case (when compared to the upper panel of Fig.~\ref{fig:cuts}).

\section{Discussion}
\label{sec:discussion}

Being able to identify collisions between GMCs is potentially
important for our understanding of the global star formation process
in galaxies.  As discussed in \S\ref{sec:bridge}, observational
surveys have revealed several cases of on-going cloud-cloud collisions
using low-$J$ CO lines as diagnostics. Our work complements these
findings by first analyzing the results of realistic MHD simulations
of colliding, turbulent, multiphase GMCs, and then exploring how
fine-structure lines compare with CO $J=1-0$. This work differs from
the studies of \citet{Hawo15a,Hawo15b} by considering a more realistic
physical model that includes magnetic fields and PDR-based
heating/cooling, a concomitant improved treatment for the abundance
and level populations of CO (whereas they assumed a constant abundance
of [CO]/[H]$=8\times10^{-5}$ throughout their clouds), and a
calculation of the radiative transfer of the fine-structure lines.

Our main finding is that the lines of [C{\sc i}] $609\,\mu{\rm m}$ and
[C{\sc ii}] $158\,\mu$m, are promising alternative ways for
identifying cloud-cloud collisions via the bridge-effect, linking
distinct velocity peaks in p-v space.  The [O{\sc i}] $63\,\mu$m line
is highly affected by the contribution of the ambient ISM, so its
emission will depend sensitively to the nature of the atomic gas
around the GMCs. Such gas has only been treated in a very idealized
way in our current simulations, so we are not able to draw strong
conclusions about the utility of [O{\sc i}] as a GMC-GMC collision
diagnostic, however, we expect it may be potentially useful if GMCs
have significant co-moving atomic halos. As the collision progresses
and the denser parts of the clouds merge, the bridge-effect diminishes
and low$-J$ CO lines then hardly show any signature of the
collision. On the other hand, the lifetime of the broad bridge in
fine-structure lines is predicted to be longer since these lines are
emitted from the more rarefied parts of GMCs which are primarily
located at their outer envelopes that have not yet undergone
collision.

The models considered here are based on an idealized situation in
which there is no local stellar feedback.  However, cloud-cloud
collisions are potential sites for triggered star formation
\citep{Fuku15,Fuku17a,Fuku17b,Tori15,Tori17}, including leading to the
birth of massive, $\gtrsim10\,{\rm M}_{\odot}$ stars
\citep{Taka14,Balf15}. Thus the signatures of the PDRs from such
localized stellar feedback \citep[see, e.g., for examples of models of
  HII region feedback][]{Dale12,Walc15,Hawo15a} may confuse those from
the bulk original GMC material, externally irradiated. Such
contributions also need to be accounted for when interpreting
observational data that includes localized PDRs, although recent
simulations by \citet{Tori17b} indicate that feedback of such massive
stars does not alter the bridge-effect in p-v diagrams of
molecular lines, such as low-$J$ $^{12}$CO. Thus the models presented
here are most useful for isolating the ``pure'' signature of the
GMC-GMC collision, which may be most easily compared to observations
of GMCs in relatively early stages of collision and star formation.

Another simplification of our treatment is that in the post-processing
of the MHD simulations, the gas is assumed to be in local
thermodynamic equilibrium and additional transient heating effects
from shocks are ignored. A future paper in this series will explore
the importance of such effects.

As discussed in \S\ref{sec:results}, most of our presented results are
for the case of the collision occurring along the line-of-sight of the
observer, which is the most idealized situation. In general, the
detectability of the bridge-effect signature will diminish as the
fraction of the collision axis that is projected along the line of
sight decreases. To overcome such effects, statistically significant
samples of GMC-GMC collision candidates need to be considered.

\section{Conclusions}
\label{sec:conclusions}

In this paper we performed synthetic observations of $^{12}$CO
$J=1-0$, [C{\sc i}] $609\,\mu{\rm m}$, [C{\sc ii}] $158\,\mu$m and
[O{\sc i}] $63\,\mu$m based on snapshots of MHD simulations of GMC
collisions, carried out in Paper II of this series. The selected
snapshots examine the colliding case of two GMCs, and the
non-colliding case of two GMCs that overlap each other along the
line-of-sight of the observer. With a simple radiative transfer tool,
we calculated integrated intensity emission maps (including of high
velocity components) and position-velocity diagrams for the above
lines.  \emph{We demonstrated that fine-structure lines, and in particular
[C{\sc ii}] $158\,\mu$m, may be a promising alternative diagnostic of
cloud-cloud collisions, separate from the more commonly discussed CO
lines.} This finding holds even when the collision has been evolved so
that the denser regions have merged and that low$-J$ CO lines are no
longer good tracers of the collision. Our results are summarized as
follows:
\begin{itemize}
\item The overall CO $J=1-0$ emission is stronger in the colliding
  than in the non-colliding case and in particular at the region where
  the collision occurs (peak of the total H-nucleus column density).
  The p-v diagram in the colliding case does not show a clear
  signature of the bridge-effect, mainly because the collision has
  been evolved to an extent that the CO-rich gas has been mixed up and
  such a signature is diminished. On the other hand, the integrated
  intensity maps of blue- and redshifted high velocity gas do show in
  a clear way whether or not the two GMCs along the line-of-sight are
  colliding.
\item The [C{\sc i}] $^3{\rm P}_1\rightarrow{^3{\rm P}}_0$ transition
  corresponding to a wavelength of $609\,\mu$m is found to be emitted
  from the more diffuse gas of both GMCs. Although it peaks at the
  position where the two GMCs collide, it does not fluctuate as much
  as the CO $J=1-0$ line and its intensity does not significantly
  differ in the non-colliding case. The bridge-effect is
  observable in the p-v diagram at $t=2\,{\rm Myr}$ in the colliding
  case, however at $t=4\,{\rm Myr}$ it has disappeared. As with the CO
  $J=1-0$ line, the high velocity maps are able to indicate the
  collision event at all times.
\item The fine-structure [C{\sc ii}] $158\,\mu$m shows the clearest
  signature of the bridge-effect in the p-v diagrams of the colliding
  case. This line is emitted from the outermost parts of both GMCs,
  with modest further contributions from the ambient ISM gas. We
    find that the contribution of the ambient ISM does not
    significantly impact the ability of [C{\sc ii}] $158\,\mu$m to be
    a tracer of GMC-GMC collisions. The overall [C{\sc ii}] emission
  does decrease in time as the collision progresses, since the
  increase in density shields the FUV radiation. The blue- and
  redshifted high velocity gas overlap almost everywhere in the
  colliding case.  
\item The emission of the [O{\sc i}] $^3{\rm P}_1\rightarrow{^3}{\rm P}_0$ transition
  at $63\,\mu$m 
is much weaker than that of [C{\sc ii}] at $158\,\mu$m, with
significant contributions from the ambient atomic ISM gas. It is thus
a better tracer of surrounding atomic halos of GMCs, which, depending
on their properties, may provide important, complementary information
to diagnose GMC-GMC collisions. 
\end{itemize}
All p-v diagrams in the non-colliding case show a velocity width
corresponding to internal turbulent gas motions, i.e., at
approximately virialized velocities. Furthermore, in all the above
cases, we have assumed that the collision occurs along the
line-of-sight of the observer. 
If the collision is viewed at an angle,
then any potential signatures will be diminished and completely
eliminated in the cases that the collision occurs mostly perpendicular
to the observer. Statistical samples of GMC-GMC collision candidate
events are needed to average over such geometric effects.

\acknowledgements  The authors thank an anonymous referee whose
  comments improved the clarity of the paper. We thank Thomas
Haworth, Adam Ginsburg and Guang-Xing Li for useful discussions. KT
thanks Kazuhito Motogi, Taishi Nakamoto and Hideko Nomura. BW is thankful for support
from the Japan Society for the Promotion of Science Postdoctoral
Fellowship.

\appendix

\section{Radiative Transfer algorithm}
\label{app} 
We describe here the algorithm of radiative transfer calculations we develop which was used in all synthetic maps in this work (see also \S\ref{ssec:RT}). We begin by solving the radiative transfer equation along the line-of-sight element $dz$:
\begin{eqnarray}
\label{eq:rteq}
\frac{dI_{\nu}}{dz}=-\alpha_{\nu}I_{\nu} + \alpha_{\nu}S_{\nu},
\end{eqnarray}
where $I_{\nu}$ is the intensity of the line, $\alpha_{\nu}$ the absorption coefficient, and $S_{\nu}$ is the source function at frequency $\nu$. The source function and the absorption coefficient for a transition of $i\rightarrow j$ are
\begin{eqnarray}
S_{\nu} = \frac{2h\nu_0^3}{c^2}\frac{n_i g_j}{n_j g_i-n_i g_j},\\
\alpha_{\nu} = \frac{c^2 n_i A_{ij}}{8\pi \nu_0^2} \left( \frac{n_j g_i}{n_i g_j}-1\right) \phi_\nu,
\end{eqnarray}
where $\nu_0$ is the frequency of the line center, $A_{ij}$ is the Einstein A coefficient, $n_i$, $n_j$ are the level populations, and $g_u$, $g_l$ are the statical weights of levels $i,~j$, and $\phi_\nu$ is the line profile. The level populations are obtained from our photodissociation region calculations using {\sc 3d-pdr} \citep[see \S\ref{ssec:3dpdr}][]{Bisb12}. We assumed a Maxwellian distribution of velocities due to the thermal and the turbulent gas motion. The line profile takes, therefore, the form
\begin{eqnarray}
\phi_{\nu}=\frac{1}{\sqrt{2\pi\sigma_{\nu}^2}}\exp\left\{-\frac{\left[\left(1+v_{\rm los}/c\right)\nu-\nu_0\right]^2}{2\sigma_{\nu}^2}\right\},
\end{eqnarray}
where $v_{\rm los}$ is the gas velocity along the line-of-sight (positive for red-shift and negative for blue-shift), and $\sigma_{\nu}$ is the dispersion defined as
\begin{eqnarray}
\label{eqn:sigma}
\sigma_{\nu}=\frac{\nu_0}{c}\sqrt{\frac{k_{\rm B}T_{\rm gas}}{m_{\rm mol}}+\frac{v_{\rm turb}^2}{2}},
\end{eqnarray}
where $T_{\rm gas}$ is the gas temperature, $m_{\rm mol}$ is the mass of the molecule/ion emitting the line, and $v_{\rm turb}$ is the a root mean-square measure of turbulent velocities (set to be $v_{\rm turb}=1.5\,{\rm km}\,{\rm s}^{-1}$ throughout the radiative transfer calculations presented here).

The formal solution of radiative transfer equation (Eqn.~\ref{eq:rteq}) is given by
\begin{eqnarray}
\label{eq:rtformal}
I_\nu(z)=I_\nu(0) e^{\tau_\nu(z)}+ \int^{\tau_\nu(z)}_0 S_\nu(z) e^{\tau'_\nu-\tau_\nu(z)}d\tau'_\nu,
\end{eqnarray}
where $\tau_\nu = \int_{0}^{z}\alpha_\nu(z')dz'$ is the optical depth.
To solve the radiative transfer numerically, the absoprtion coefficient is assumed to be a linear function of $z$ and the source function is assumed to be a linear function of $\tau$ between each grid  $[z_{p},~z_{p+1}]$ \citep[see also][for alternative ways of solving the radiative transfer equation in this limit]{Andr17}.
Then, the differential solution can be obtained as
\begin{eqnarray}
\label{eq:rtformal2}
I_{\nu,p+1}=I_{\nu,p}e^{-\Delta\tau_{\nu}}+S_{\nu,p}\left(\frac{1-e^{-\Delta\tau_{\nu}}}{\Delta\tau_{\nu}}-e^{\Delta\tau_{\nu}}\right)+S_{\nu,p+1}\left(1-\frac{1-e^{-\Delta\tau_{\nu}}}{\Delta\tau_{\nu}}\right),\\
\Delta \tau_\nu = \frac{\alpha_{\nu,p}+\alpha_{\nu,p+1}}{2}dz,
\end{eqnarray}
where the subscripts of $p$ and $p+1$ denote the position of variables. This form of the solution can reproduce the physical consequences at both optically thin and thick limits:
\begin{eqnarray}
I_{\nu,p+1}=
\begin{cases}
I_{\nu,p}(1-\Delta\tau_{\nu})+\frac{B_{\nu,p}+B_{\nu,p+1}}{2}\Delta\tau_{\nu} & (\Delta\tau_\nu \rightarrow 0),\\
B_{\nu,p+1} & (\Delta\tau_\nu \rightarrow \infty).
\end{cases}
\end{eqnarray}
For boundary conditions, we adopt the intensity of the Cosmic Microwave Background Radiation of $2.7\,{\rm K}$, i.e. $I_\nu(z=0)=B_\nu(2.7K)$ where $B_\nu$ is the Planck function. The line profile is obtained by solving Eqn.~\ref{eq:rtformal} for the frequency range $[\nu_0-\Delta\nu/2,\,\nu_0+\Delta\nu/2]$, where $\Delta\nu\ll\nu_0$ is the band width.

\end{document}